# QUANTUM TURING MACHINES: LOCAL TRANSITION, PREPARATION, MEASUREMENT, AND HALTING

Masanao Ozawa

School of Informatics and Sciences
Nagoya University
Chikusa-ku, Nagoya 4648601, Japan

Foundations of the theory of quantum Turing machines are investigated. The protocol for the preparation and the measurement of quantum Turing machines is discussed. The local transition functions are characterized for fully general quantum Turing machines. A new halting protocol is proposed without augmenting the halting qubit and is shown to work without spoiling the computation.

## INTRODUCTION

The Church-Turing thesis[1, 2] states that to be computable is to be computable by a Turing machine and the modern discipline in computational complexity theory[3] states that to be efficiently computable is to be computable by a Turing machine within polynomial steps in the length of the input data. However, Feynman[4] pointed out that a Turing machine cannot simulate a quantum mechanical process efficiently and suggested that a computing machine based on quantum mechanics might be more powerful than Turing machines. Deutsch introduced quantum Turing machines[5] and quantum circuits[6] for establishing the notion of quantum algorithm exploiting "quantum parallelism". A different approach to quantum Turing machines was investigated earlier by Benioff[7]. Bernstein and Vazirani[8] instituted quantum complexity theory based on quantum Turing machines and showed constructions of universal quantum Turing machines. Yao[9] showed that a computation by a quantum Turing machine can be simulated efficiently by a quantum circuit. Deutsch's idea of quantum parallelism was realized strikingly by Shor[10], who found efficient quantum algorithms for the factoring problem and the discrete logarithm problem, for which no efficient algorithms have been found for classical computing machines. The purpose of this paper is to discuss foundations of quantum Turing machines and to propose a computational protocol for quantum Turing machines.

A precise formulation of quantum Turing machines is given along with Deutsch's formulation[5] and the computational protocol is discussed for the preparation and the measurement of quantum Turing machines.

The characterization of the transition functions of quantum Turing machines is also discussed. Deutsch[5] required that the transition function should be determined by local configurations. Bernstein and Vazirani[8] found a simple characterization of the transition functions for the restricted class of quantum Turing machines in which the head must move either to the right or to the left at each step but a general characterization remains open. This problem is discussed and a solution is given.

The computational protocol for the halting of quantum Turing machines is discussed. In order to signal the completion of computation, Deutsch[5] introduced the halt flag by augmenting the halt qubit. Myers[11] pointed out a difficulty in this halting of quantum Turing machines. With improving the preceding work[12], a halting protocol is proposed without augmenting the halting qubit and it is shown that the monitoring of the halt flag does not spoil the computation.

## QUANTUM TURING MACHINES

A *quantum Turing machine (QTM)* $\mathcal{Q}$ is a quantum system consisting of a *processor*, a bilateral infinite *tape*, and a *head* to read and write a symbol on the tape. Its configuration is determined by the *processor configuration* $q$ from a finite set $Q$ of symbols, the *tape configuration* $T$ represented by an infinite string from a finite set $\Sigma$ of symbols, and the discretized *head position* $\xi$, taking values in the set $\mathbf{Z}$ of integers. The tape consists of *cells* numbered by the integers and the head position $\xi$ is the place of the cell numbered by $\xi$. We assume that $Q$ contains two specific symbols $q_0$ and $q_f$ representing the initial configuration and the final configuration of the processor and that $\Sigma$ contains the symbol $B$ representing the blank cell in the tape. For any integer $m$ the symbol at the cell $m$ on the tape is denoted by $T(m)$. We assume that the possible tape configurations are such that $T(m) = B$ except for finitely many cells $m$. The set of all the possible tape configurations is denoted by $\Sigma^{\#}$. The set $\Sigma^{\#}$ is a countable set. Thus, any configuration $C$ of $\mathcal{Q}$ is represented by a triple $C = (q, T, \xi)$ in the *configuration space* $Q \times \Sigma^{\#} \times \mathbf{Z}$. The state of $\mathcal{Q}$ is represented by a unit vector in the Hilbert space $\mathcal{H}$ generated by the configuration space $Q \times \Sigma^{\#} \times \mathbf{Z}$. The complete orthonormal basis canonically in one-to-one correspondence with the configuration space is called the *computational basis*. Thus, the computational basis is represented by $|C\rangle = |q\rangle|T\rangle|\xi\rangle$ for any configuration $C = (q, T, \xi) \in Q \times \Sigma^{\#} \times \mathbf{Z}$; we shall write also $|q, T, \xi\rangle = |q\rangle|T\rangle|\xi\rangle$.

We shall denote by $|X|$ the number of the elements of a set $X$; for an indexed set the number of elements is understood as the number of indices. In order to define the observables quantizing the configurations, we assume the numbering of the sets $Q$ and $\Sigma$ such that $Q = \{q_0, \ldots, q_{|Q|-1}\}$ and $\Sigma = \{\sigma_0, \ldots, \sigma_{|\Sigma|-1}\}$. We define observables $\hat{q}$, $\hat{T}(m)$ for $m \in \mathbf{Z}$, and $\hat{\xi}$ as follows.

$$\hat{q} = \sum_{n=0}^{|Q|-1} n|q_n\rangle\langle q_n|, \quad \hat{T}(m) = \sum_{n=0}^{|\Sigma|-1} n|\sigma_n\rangle\langle \sigma_n|, \quad \hat{\xi} = \sum_{\xi \in \mathbf{Z}} \xi|\xi\rangle\langle \xi|.$$

We assume that we have a device to prepare the quantum Turing machine in the state $|q, T, \xi\rangle$ for any configuration $C = (q, T, \xi)$ and that we have a measuring device to measure sufficiently many $\hat{T}(m)$s simultaneously.

Let $\Gamma$ be a finite set of symbols and $\Gamma^*$ the set of finite strings from $\Gamma$. In this paper, we shall consider computations which are probabilistic transformations on $\Gamma^*$, or precisely functions from $\Gamma^*$ to the set of probability distributions on $\Gamma^*$. The set $\Gamma$ is called the *alphabet* of the computation. A finite string from the set $\Gamma$ is called a $\Gamma$-*string*. The length of a $\Gamma$-string $x$ is denoted by $|x|$. When $|x| = 0$, $x$ is called the empty string. We shall identify any $\Gamma$-string $x = (x_0, \ldots, x_{|x|-1})$ with a function $x$ from $\{0, \ldots, |x|-1\}$ to $\Gamma$ such that $x(m) = x_m$ for all $m$ with $0 \leq m \leq |x|-1$.

The computation by a QTM consists of encoding, preparation, time evolution, measurement, and decoding. The encoding transforms the *input Γ-string* to the *input tape string*. The preparation prepares the *initial state* of the quantum Turing machine with the input tape string, and the time evolution transforms the initial state to the *final state*. The measurement of the tape string in the final state gives a probability distribution of the *output tape string*. The decoding transforms the output tape string to the output Γ-string and hence transforms the probability distribution of the output tape string to the probability distribution of the output Γ-string. Therefore, the initial Γ-string is transformed to the *output probability distribution* of the Γ-string.

The *encoding e* of the QTM $\mathcal{Q}$ is a polynomial time computable function from $\Gamma^*$ to $\Sigma^\#$. Thus, the encoding $e$ transforms any Γ-string $x$ to a tape configuration denoted by $e(x)$; if $T = e(x)$ we shall write $T \sim x$ and $T$ is said to *represent* the Γ-string $x$. Inversely, the *decoding d* of $\mathcal{Q}$ is a polynomial time computable function from $\Sigma^\#$ to $\Gamma^*$ satisfying $d(e(x)) = x$ for all $x \in \Gamma^*$.

In this paper, we assume that $B \notin \Gamma$ and $\Sigma = \Gamma \cup \{B\}$. We assume that there is an infinite subset $S \subset \mathbf{N}$ of the set of tape cells, called the *data slot*, with polynomial time numbering $S = \{m_1, m_2, \ldots\}$ and that the encoding is such that

$$e(x)(m) = \begin{cases} x(n) & \text{if } m = m_n \in S \text{ and } 0 \leq n < |x|, \\ B & \text{otherwise,} \end{cases} \quad (1)$$

for any $x \in \Sigma^*$, and the decoding is given by

$$|d(T)| = \min\{m_n \in S|\ T(m_n) = B\}, \quad (2)$$
$$d(T)(n) = T(m_n) \quad (3)$$

for $0 \leq n < |d(T)|$, where $T \in \Sigma^\#$.

The computation begins at $t = 0$. At this time $\mathcal{Q}$ is prepared in an *initial state* $|C_0\rangle$ such that

$$|C_0\rangle = |q_0\rangle|T_{in}\rangle|0\rangle, \quad (4)$$

where $T_{in}$ represents some Γ-string $x$. In this case, $T_{in}$ is called the *input tape*, $x$ is called the *input*, and $|x|$ is called the *input length*.

The computation proceeds in steps of a fixed unit duration $\tau$. Since the position of the head is discretized, the wave function $|\psi(t)\rangle$ may not stay within $\mathcal{H}$ at any time $t$ other than integer multiples of $\tau$. We assume therefore that the time $t$ is discretized to be an integer multiple of $\tau$. We also take the normalized unit of time in which the time $t$ is assumed to take values in $\mathbf{Z}$. The dynamics of $\mathcal{Q}$ are described by a unitary operator $U$ on $\mathcal{H}$ which specifies the evolution of any state $|\psi(t)\rangle$ during a single *computational step* so that we have

$$U^\dagger U = UU^\dagger = I, \quad (5)$$
$$|\psi(t)\rangle = U^t|\psi(0)\rangle \quad (6)$$

for all positive integer $t$.

Since the number of all the possible tape strings in the data slot is countable, we assume them to be indexed as $\{T_1, T_2, \ldots\}$. Thus, the observable $\hat{T}(S)$ describing the tape string in the data slot can be represented by

$$\hat{T}(S) = \sum_{j=1}^{\infty} \lambda_j\ I_1 \otimes |T_j\rangle\langle T_j| \otimes I_2 \otimes I_3$$

where $\{\lambda_1, \lambda_2, \ldots\}$ is a countable set of positive numbers in one-to-one correspondence with $\{T_1, T_2, \ldots\}$ by a polynomial time function and where $I_1$ is the identity on the state space spanned by the processor configurations $Q$, $I_2$ is the identity on the state space spanned by the tape strings outside the data slot, and $I_3$ is the identity on the state space spanned by the head positions $\mathbf{Z}$.

We assume that the measurement to obtain the output is allowed only for the computational basis or more specifically the observable $\hat{T}(S)$ describing directly the output symbol string on the tape, while in Deutsch's formulation[5,6] and in later work no such restriction has been taken place. However, it is an unavoidable assumption in the definition of quantum Turing machine. In fact, if this assumption is dropped, any function would computable without any computational time. To see this, suppose that the tape strings are encoded by the natural numbers. Let $|T_n\rangle$ be the computational basis state, ignoring the inessential degeneracy, in which the output tape string is the one encoded by $n$ and let $\hat{T}$ be the observable such that $\hat{T}|n\rangle = n|T_n\rangle$. Only such $\hat{T}$ is allowed to measure for obtaining the output. Otherwise, given any function $f$ of the natural numbers and a natural number $n$, if one prepares the tape in the state $|T_n\rangle$ and measures the observable $f(\hat{T})$, one gets $f(n)$ surely without any computation. This contradicts the Church-Turing thesis. Thus, we cannot allow even the measurement of $f(\hat{T})$ unless $f$ is a polynomial time computable function.

## LOCAL TRANSITION FUNCTIONS

Deutsch[5] requires that the QTM operate finitely, i.e., (i) only a finite system is in motion during any one step, (ii) the motion depends only on the state of a finite subsystem, and (iii) the rule that specifies the motion can be given finitely in the mathematical sense. To satisfy the above requirement, the matrix elements of $U$ takes the following form*:

$$\langle q', T', \xi'|U|q, T, \xi\rangle = [\delta_{\xi'}^{\xi+1} D(q, T(\xi), q', T'(\xi), 1) + \delta_{\xi'}^{\xi} D(q, T(\xi), q', T'(\xi), 0) \\ + \delta_{\xi'}^{\xi-1} D(q, T(\xi), q', T'(\xi), -1)] \prod_{m \neq \xi} \delta_{T(m)}^{T'(m)} \quad (7)$$

for any configurations $(q, T, \xi)$ and $(q', T', \xi')$. The continued product on the right ensures that the tape is changed only at the head position $\xi$ at the beginning of each computational step. The terms $\delta_{\xi'}^{\xi\pm 1}$, $\delta_{\xi'}^{\xi}$ ensure that during each step the head position cannot change by more than one unit. The function $D(q, T(\xi), q', T'(\xi), d)$, where $q, q' \in Q$, $T(\xi), T'(\xi) \in \Sigma$, and $d \in \{-1, 0, 1\}$, represents a dynamical motion depending only on the local observables $\hat{q}$ and $\hat{T}(\xi)$. We call $D$ the *local transition function* of the QTM $\mathcal{Q}$.

The function $D$ can be arbitrarily given except for the requirement (5) that $U$ be unitary. Each choice defines a different QTM $\mathcal{Q}[D]$. Thus, if we have an intrinsic characterization of the local transition function $D$, QTMs can be defined formally without referring to the unitary operator $U$ as a primitive notion.

From (7), the time evolution operator $U$ is determined conversely from the local transition function $D$

$$U|q, T, \xi\rangle = \sum_{p, \tau, d} D(q, T(\xi), p, \tau, d)|p, T_\xi^\tau, \xi + d\rangle. \quad (8)$$

---

*This condition is a natural extension of Deutsch's condition[5] to the case where the head is not required to move.

for any configuration $(q, T, \xi)$, where $T_\xi^\tau$ is the tape string defined by

$$T_\xi^\tau(m) = \begin{cases} \tau & \text{if } m = \xi, \\ T(m) & \text{if } m \neq \xi. \end{cases} \quad (9)$$

It follows that the relation $D(q, \sigma, q', \tau, d) = c$ can be interpreted as the following instruction of the operation of $\mathcal{Q}$: if the processor is in the configuration $q$ and if the head reads the symbol $\sigma$, then it follows with amplitudes $c$ that the processor's state turns to $q'$, the head writes the symbol $\tau$, and that the head moves one cell to the right if $d = 1$, to the left if $d = -1$, or does not move if $d = 0$.

Now we can formulate the characterization problem of local transition functions of QTMs: *Let $D$ be a complex-valued function on $Q \times \Sigma \times Q \times \Sigma \times \{-1, 0, 1\}$ and let $U$ be the operator on $\mathcal{H}$ defined by (8). Then, what conditions ensure that the operator $U$ is unitary?*

This problem is solved by the following theorem.[13]

**Theorem 1** *The operator $U$ is unitary if and only if $D$ satisfies the following conditions.*

(a) *For any $(q, \sigma) \in Q \times \Sigma$,*

$$\sum_{p,\tau,d} |D(q, \sigma, p, \tau, d)|^2 = 1.$$

(b) *For any $(q, \sigma), (q', \sigma') \in Q \times \Sigma$ with $(q, \sigma) \neq (q', \sigma')$,*

$$\sum_{p,\tau,d} D(q', \sigma', p, \tau, d)^* D(q, \sigma, p, \tau, d) = 0.$$

(c) *For any $(q, \sigma, \tau), (q', \sigma', \tau') \in Q \times \Sigma^2$, we have*

$$\sum_{p \in Q} D(q', \sigma', p, \tau', 1)^* D(q, \sigma, p, \tau, -1) = 0.$$

(d) *For any $(q, \sigma, \tau), (q', \sigma', \tau') \in Q \times \Sigma^2$, we have*

$$\sum_{p \in Q, d=0,1} D(q', \sigma', p, \tau', d-1)^* D(q, \sigma, p, \tau, d) = 0.$$

If it is assumed that the head must move either to the right or to the left at each step, the condition (d) is automatically satisfied. In this case, the above statement is reduced to the result due to Bernstein and Vazirani[8].

In order to maintain the Church-Turing thesis, we need to require that the unitary operator $U$ is constructive, or that the matrix elements of $U$ in the computational basis are computable complex numbers; otherwise, we cannot show the existence of the algorithm by the constructive language. From the complexity theoretical point of view, we need also to require that matrix elements are polynomially computable complex numbers. Thus, we require that the range of the transition function $\delta$ is in the polynomially computable complex numbers.

## HALTING PROTOCOL

The result of a computation is obtained by measuring the tape string after the computation has been completed. Unlike the classical case, the machine configuration cannot be monitored throughout the computation because of the inevitable disturbance caused by measurement. Thus, the machine needs a specific halt scheme to signal actively when the computation has been completed.

Deutsch[5] introduced an additional single qubit, called the halt qubit, together with an observable $\hat{n}_0$, called the halt flag, with the eigenstates $|0\rangle$ and $|1\rangle$, so that the processor configuration $q$ is represented by the state vector $|q\rangle|1\rangle$ if $q$ is the final state in the classical picture or by $|q\rangle|0\rangle$ otherwise. The halt qubit is initialized to $|0\rangle$ before starting the computation, and every valid quantum algorithm sets the halt qubit to $|1\rangle$ when the computation has been completed but does not interact with the halt qubit otherwise. Deutsch claimed that *the observable $\hat{n}_0$ can then be periodically observed from the outside without affecting the operation of the machine.*

Myers[11] argued that the state entangles the non-halt qubits with the halt qubits so that the measurement of the halt flag changes the state and concluded that the halt scheme spoils the computation.

In the preceding work[12], Deutsch's halt scheme is reformulated precisely and it is shown that, even though it changes the state of the quantum Turing machine, the measurement of the halt flag does not change the probability distribution of the outcome of the computation so that it does not spoil the computation. It is also shown[12] that the halt scheme is equivalent to the quantum nondemolition monitoring of the output observable.

In what follows, we shall give a new formulation of the halt scheme in which the additional halt qubit is not augmented.

The *halt flag* $\hat{n}_0$ is defined to be the observable corresponding to the projection on the final configuration of the processor, i.e.

$$\hat{n}_0 = |q_f\rangle\langle q_f|. \quad (10)$$

We assume that we have a measuring apparatus to measure $\hat{n}_0$ precisely after each step instantaneously in the manner satisfying the projection postulate. Thus the $\hat{n}_0$-measurement gives surely the outcome 1 if and only if the processor is in $|q_f\rangle$. We shall denote by $[\![\hat{A} = a]\!]$ the spectral projection onto the eigenspace of an observable $\hat{A}$, considered as an operator on $\mathcal{H}$, corresponding to the eigenvalue $a$. The product $[\![\hat{A} = a]\!][\![\hat{B} = b]\!]$, if commutable, will be denoted by $[\![\hat{A} = a, \hat{B} = b]\!]$.

The precise formulation of the halting protocol is given as follows.

(I) The halt flag $\hat{n}_0$ is measured instantaneously after every step. This measurement is a precise measurement of the observable $\hat{n}_0$ satisfying the projection postulate. (Note that the above measurement is different from the procedure that one measures $\hat{q}$ and checks if the outcome is $q_f$ because this does not satisfy the projection postulate.)

(II) Once the halt flag is set to $\hat{n}_0 = 1$, the QTM no more changes the halt flag nor the result of computation. Thus, we require

$$U[\![\hat{n}_0 = 1, \hat{T}(S) = T_j]\!]U^t|C\rangle = [\![\hat{n}_0 = 1, \hat{T}(S) = T_j]\!]U[\![\hat{n}_0 = 1, \hat{T}(S) = T_j]\!]U^t|C\rangle \quad (11)$$

for any initial configuration $C$, time $t \geq 0$, and tape string $T_j$ over the data slot $S$.

(III) After the measurement of the halt flag $\hat{n}_0$ gives the outcome 1, the tape string $\hat{T}(S)$ in the date slot is measured and the outcome of this measurement is defined to be the *output* of the computation.

Now we shall show that the halting protocol does not affect the result of the computation. For that purpose, it suffices to prove that the probability distribution of the output is not affected by monitoring of the halt flag.

Let $\Pr\{\text{output} = T_j|\text{monitored}\}$ be the probability of finding the output $T_j$ up to $N$ steps by the halting protocol. Let $\Pr\{\text{output} = T_j|\text{not-monitored}\}$ be the probability of finding the output $T_j$ by the single measurement after $N$ steps. We shall prove

$$\Pr\{\text{output} = T_j|\text{monitored}\} = \Pr\{\text{output} = T_j|\text{not-monitored}\} \tag{12}$$

Let $P = [\![\hat{n}_0 = 1]\!]$ and $Q_j = [\![\hat{T}(S) = T_j]\!]$. Let $|C\rangle$ be an arbitrary initial state. If $|C\rangle$ is the state of the machine before the computation, We have

$$\Pr\{\text{output} = T_j|\text{not-monitored}\} = \|PQ_j U^N |C\rangle\|^2. \tag{13}$$

By the projection postulate, the joint probability of obtaining the outcome $\hat{n}_0 = 0$ at the times $1, \ldots, K-1$ and obtaining the outcomes $\hat{n}_0 = 1$ and $\hat{T} = \lambda_j$ at the time $K$ is given by

$$\|PQ_j(UP^\perp)^K|C\rangle\|^2, \tag{14}$$

and hence we have

$$\Pr\{\text{output} = T_j|\text{monitored}\}$$
$$= \|PQ_j|C\rangle\|^2 + \|PQ_j UP^\perp|C\rangle\|^2 + \cdots + \|PQ_j(UP^\perp)^N|C\rangle\|^2. \tag{15}$$

Thus, it suffices to prove the relation

$$\|PQ_j U^N |C\rangle\|^2 = \|PQ_j|C\rangle\|^2 + \|PQ_j UP^\perp|C\rangle\|^2 + \cdots + \|PQ_j(UP^\perp)^N|C\rangle\|^2 \tag{16}$$

for any $N$ and any initial state $|C\rangle$.

Let $\psi = U^t|C\rangle$ where $t \geq 0$. We first consider the relation

$$\|PQ_j U\psi\|^2 = \|PQ_j\psi\|^2 + \|PQ_j UP^\perp \psi\|^2. \tag{17}$$

From (11), we have

$$PQ_j UPQ_j \psi = UPQ_j \psi. \tag{18}$$

It follows that

$$PQ_j UPQ_j^\perp \psi = \sum_{k \neq j} PQ_j UPQ_k \psi = 0. \tag{19}$$

From (18) and (19), we have

$$PQ_j U\psi = PQ_j UPQ_j \psi + PQ_j UPQ_j^\perp \psi + PQ_j UP^\perp \psi$$
$$= UPQ_j \psi + PQ_j UP^\perp \psi. \tag{20}$$

From (18), we have

$$\langle UPQ_j \psi | PQ_j UP^\perp \psi \rangle = \langle PQ_j UPQ_j \psi | UP^\perp \psi \rangle$$
$$= \langle UPQ_j \psi | UP^\perp \psi \rangle$$
$$= 0, \tag{21}$$

From (20) and (21), we have

$$\begin{aligned}\|PQ_jU\psi\|^2 &= \|UPQ_j\psi + PQ_jUP^\perp\psi\|^2\\ &= \|UPQ_j\psi\|^2 + \|PQ_jUP^\perp\psi\|^2\\ &= \|PQ_j\psi\|^2 + \|PQ_jUP^\perp\psi\|^2\end{aligned} \qquad (22)$$

Thus, we have proved (17).

The proof for general $N$ runs as follows. We use mathematical induction and assume that (16) holds for $N-1$. By replacing $\psi$ by $U^{N-1}|C\rangle$ in (17), we have

$$\|PQ_jU^N|C\rangle\|^2 = \|PQ_jU^{N-1}|C\rangle\|^2 + \|PQ_jUP^\perp U^{N-1}|C\rangle\|^2. \qquad (23)$$

From (18), we have $P^\perp UP\psi = \sum_j P^\perp UPQ_j\psi = 0$, and hence $P^\perp U\psi = P^\perp UP^\perp\psi$ so that $P^\perp U^{N-1}|C\rangle = P^\perp(UP^\perp)^{N-1}|C\rangle$. It follows that

$$\|PQ_jUP^\perp U^{N-1}|C\rangle\|^2 = \|PQ_j(UP^\perp)^N|C\rangle\|^2. \qquad (24)$$

By induction hypothesis, we have

$$\|PQ_jU^{N-1}|C\rangle\|^2 = \|PQ_j|C\rangle\|^2 + \|PQ_jUP^\perp|C\rangle\|^2 + \cdots + \|PQ_j(UP^\perp)^{N-1}|C\rangle\|^2. \qquad (25)$$

Therefore, from (23), (24), and (25), we obtain (16).

It is concluded that the the probability of finding the output $T_j$ up to $N$ steps by the halt protocol is equal to the probability of finding the output $T_i$ by the single measurement of $\hat{T}(S)$ after $N$ steps. It follows that the halting protocol does not affect the result of the computation.

Recently, Linden and Popescu[14] claimed that the halt scheme given previously[12] is not consistent with unitarity of the evolution operator. However, their argument applies only to the special case in which the whole tape is required not to change after the halt. As suggested in a footnote, the conclusion in the previous work[12] can be obtained from the weaker condition for the general case where the tape is allowed to change except for the date slot. Linden and Popescu[14] disregarded this case and hence their conclusion is not generally true. In this paper, the halting protocol with such a general formulation is treated explicitly and it is proved that even in this case the computation is not affected by the measurement of the halt flag. Moreover, contrary to Linden and Popescu[14], this general formulation is consistent with the unitarity. In fact, it can be shown that any unidirectional QTMs and stationary QTMs[8] can be simulated by QTMs obeying this halting protocol with constant slowdown[13]. Thus, there is a universal QTM obeying the halting protocol.